\documentclass[11pt]{article}
\pdfoutput=1
\usepackage{graphics}
\usepackage{latexsym}
\newcommand{\xxx}[1]{}
\newcommand{\mysection}[1]{\section{#1}
   \hspace{0.8cm}\setcounter{equation}{0}}

\newenvironment{MYitemize}[1]{\renewcommand{\labelitemi}{#1}
\begin{itemize}}{\end{itemize}}
\DeclareMathAlphabet   {\mathsc}{OT1}{cmr}{m}{sc}

\newcommand{\myref}[1]{(\ref{#1})}
\newcommand{\mysec}[1]{Section~\ref{#1}}

\def\Del{\Delta}
\def\half{{1\over2}}
\def\bea{\begin{eqnarray}}
\def\eea{\end{eqnarray}}
\def\beq{\begin{equation}}
\def\eeq{\end{equation}}

\def\ux{$U(1)_X$}

\def\uo{$U(1)$}

\def\vx{V_X}

\def\superint{\int d^{4}\theta}
\def\M{\bar{M}}

\newcommand{\WaWa}{ W_a^{\alpha} W^a_{\alpha}}

\newcommand{\Wa}{W_{\alpha}}

\newcommand{\Wc}{W^{\alpha}}

\def\vev{$vev$}
\def\re{{\rm Re}}
\def\im{{\rm Im}}

\def\bM{\bar{M}}

\def\W{\overline{W}}

\def\tF{{\tilde F}}

\def\D{{\cal D}}

\def\bD{\bar{\D}}

\def\pp{\partial}

\def\notD{\not{\hspace{-.05in}D}}

\def\notcD{\not{\hspace{-.05in}\D}}

\def\ibar{\bar{\imath}}

\def\[{\left [}
\def\]{\right ]}
\def\({\left (}
\def\){\right )}

\def\r{\right|}
\def\l{\left.}

\def\H{\bar{H}}

\def\T{\bar{T}}

\def\z{\bar{z}}

\def\del{\delta}

\def\S{{\bar{S}}}

\def\STr{{\rm STr}}
\def\Tr{{\rm Tr}}

\def\Ti^{T^{(i)}}

\def\f{\bar{f}}

\def\L{{\cal L}}

\def\hz{\hat{z}}

\def\hA{\hat{A}}

\def\n{\bar{n}}
\def\m{\bar{m}}

\def\A{\bar{A}}

\def\Z{{\bar{Z}}}

\def\bF{\bar{F}}

\def\a0{\alpha_0}
\def\chiproj{(\bD^2 - 8R)}

\def\eee{\nonumber \\ &=&}
\def\ddd{\nonumber \\ &&}

\def\nnn{\nonumber \\ }
\def\hc{ + {\rm h.c.}}

\def\thefootnote{\fnsymbol{footnote}}

\newcommand{\myvev}[1]{{\langle #1 \rangle}}
\newcommand{\gappeq}{\mathrel{\rlap {\raise.5ex\hbox{$>$}}
{\lower.5ex\hbox{$\sim$}}}}
\newcommand{\lappeq}{\mathrel{\rlap{\raise.5ex\hbox{$<$}}
{\lower.5ex\hbox{$\sim$}}}}

\def\ep{\epsilon}

\begin{document}
\begin{titlepage}

\hfill   LBNL-

\hfill   UCB-PTH0-11/07

%\hfill   hep-th/

\hfill   September 2011
%\hfill \today

\begin{center}

\vspace{18pt}
{\bf BRST Invariant PV Regularization of SUSY Yang-Mills and 
SUGRA}\footnote{This work was supported in part by the
Director, Office of Science, Office of High Energy and Nuclear
Physics, Division of High Energy Physics of the U.S. Department of
Energy under Contract DE-AC02-05CH11231, in part by the National
Science Foundation under grant PHY-0457315.}\footnote{To be published
in the proceedings of ``Raymond's 80th Birthday Party'',
LAPP, Annecy-le-Vieux, July 8, 2011.}

\vspace{18pt}

Mary K. Gaillard% {\em and} Bruno Zumino
\vskip .01in
{\em Department of Physics, University of California 
and \\ Theoretical Physics Group, Bldg. 50A5104,
Lawrence Berkeley National Laboratory \\ Berkeley,
CA 94720 USA}

\vspace{18pt}

\end{center}

\begin{abstract} Pauli-Villars regularization of Yang-Mills theories
  and of supergravity theories is outlined, with an emphasis on BRST
  invariance. Applications to phenomenology and the anomaly structure
  of supergravity are discussed.

\end{abstract}

\end{titlepage}

\newpage
%\null

%
\renewcommand{\thepage}{\roman{page}}
\setcounter{page}{2}
\mbox{ }

\vskip 1in

\begin{center}
{\bf Disclaimer}
\end{center}

\vskip .2in

\begin{scriptsize}
\begin{quotation}
This document was prepared as an account of work sponsored by the United
States Government. Neither the United States Government nor any agency
thereof, nor The Regents of the University of California, nor any of their
employees, makes any warranty, express or implied, or assumes any legal
liability or responsibility for the accuracy, completeness, or usefulness
of any information, apparatus, product, or process disclosed, or represents
that its use would not infringe privately owned rights. Reference herein
to any specific commercial products process, or service by its trade name,
trademark, manufacturer, or otherwise, does not necessarily constitute or
imply its endorsement, recommendation, or favoring by the United States
Government or any agency thereof, or The Regents of the University of
California. The views and opinions of authors expressed herein do not
necessarily state or reflect those of the United States Government or any
agency thereof of The Regents of the University of California and shall
not be used for advertising or product endorsement purposes.
\end{quotation}
\end{scriptsize}

\vskip 2in

\begin{center}
\begin{small}
{\it Lawrence Berkeley Laboratory is an equal opportunity employer.}
\end{small}
\end{center}

\newpage
\renewcommand{\thepage}{\roman{page}}

\begin{center}\section*{BRST INVARIANT PV REGULARIZATION OF SUSY
  YANG-MILLS AND SUGRA}

Mary K. Gaillard\\
University of California at Berkeley\end{center}

\renewcommand{\thepage}{\arabic{page}}
\setcounter{page}{1}

\def\thefootnote{\arabic{footnote}} \setcounter{footnote}{0}
%\mysection{Introduction} 
%
Since I never collaborated on a paper with Raymond, I chose a topic
that at least allowed me to put his initial in the title.  I have been
working for a number of years on Pauli-Villars (PV) regularization of
supersymmetric theories and its applications, and I often get the
question ``Aren't you breaking BRST~\cite{brst} invariance?'' In the
following I will explain how the miraculous cancellations among boson and
fermion loops in supersymmetry (SUSY) allow for the complete
elimination of ultraviolet (UV) divergences by the introduction of
only chiral supermultiplets and, in the case of local supersymmetry
(supergravity or SUGRA) Abelian gauge supermultiplets.  I will also
describe some applications to phenomenology, and will discuss
conformal and chiral anomalies in supergravity, and their cancellation
in the context of effective theories from compactification of the
weakly coupled heterotic string (WCHS).

\mysection{SUSY Yang-Mills with chiral matter}\label{susyym}
A renormalizable, globally supersymmetric, theory is defined by two
types of chiral superfields: $Z^i$ for matter, with components 
$(z^i,\chi^i_\alpha,F^i)$, and the Yang-Mills (YM) superfield strengths 
$\Wa^a$ with components $(\lambda^a_\alpha,F^a_{\mu\nu},D^a)$, where 
$\alpha$ is a Dirac index, $i,a$ denote internal quantum numbers, 
and $F^i,D^a$ are auxiliary fields.  The theory is further defined by
the superpotential: 
\beq W(Z)= \half\mu_{i j}Z^i Z^j + {1\over6}c_{i j k}Z^i Z^j Z^k,
\label{superpot}\eeq
and gauge transformation properties of the matter supermultiplets: 
\beq \del_a Z^i = i(T_a Z)^i,\qquad\del_a Z^{\m} = -i(T_a^T Z)^{\m}.
\label{charges}\eeq
There are no quadratic UV divergences; these are determined by the
supertrace of the (field-dependent) mass matrix\footnote{Throughout I
  use background field techniques and set fermions to zero in the
  background; the one-loop effective fermionic Lagrangian can be
  inferred from the bosonic result by supersymmetry.} which vanishes
identically in this theory:
\beq \STr M^2(z^i,\z^{\m},F_{\mu\nu}^a)=\sum_{S= 0,\half,1}(-1)^{2S}(2S+1)M^2_S = 0\eeq
The only logarithmic UV divergences are in wave function
renormalizations.  In particular the $\beta$-function is proportional
to the parameter
\beq b_a = - {1\over16\pi^2}(3C_a - C_a^M) = g^{-3}(\mu)
{\pp g_a(\mu)\over\pp\ln\mu} = g^{-2}_a(\mu)\beta_a(\mu),\label{beta}\eeq

where $C_a$ and $C_a^M$ are quadratic Casimirs in the adjoint and
matter representations, respectively.  The superpotential
\myref{superpot} is not renormalized; with a ``supersymmetric"
choice~\cite{bfmps} of gauge fixing (to be made explicit in the next
section), the UV divergent contribution to the scalar potential is
\beq \Del V = {1\over64\pi^2}\STr M^4\ln\Lambda^2 = 
- \half\sum_a b_a D_a^2\ln\Lambda^2,\eeq
which is just the supersymmetric completion of the vacuum polarization.
In this gauge the anomalous dimension (matrix) for chiral superfields
$Z^i$ is given by
\beq 32\pi^2\gamma_i^j = -4g^2\sum_a C^a_2(r^i)\del_i^j +
\sum_{k l}c_{k l i}\bar c^{k l j}, \qquad
C^a_2(r) = {{\rm dim}\,a\over{\rm dim}\,r}C_a^r\label{gamma}.\eeq
The logarithmic divergences of this theory can be
canceled~\cite{susypv} by adding chiral PV supermultiplets
$Z^I,Y_I,\varphi^a$ with gauge transformation properties
\beq \del^a Z^I = i(T^a Z)^I,\qquad\del_a Y_I = -i(Y T_a)_I,\qquad
\del^a \varphi^b = f^{a b c} \varphi_c,\qquad (+ 
{\;\rm other\; reps}),\label{pvcharges}\eeq
where $f^{a b c}$ is a structure constant of the gauge group,
and superpotential couplings
\beq W_{\rm P V} = \half\(\mu_{i j} + c_{i j k}Z^k\)Z^I Z^J + 
\sqrt{2}g\varphi^a(T_a Z)^iY_I,\eeq
leaving BRST unbroken. Cancellation of (one-loop) UV divergences is
assured {\it provided} 
\beq C^M_a = \(\Tr T_a^2\)_{\rm matter} = \Tr(T^R_a)^2\label{gcond}\eeq 
for some real representation $R$ because one has to give gauge
invariant masses to all the PV fields, and therefore they must form an
overall real (reducible) representation which cancels the matter
contribution to \myref{beta} -- hence the ``other reps" in
\myref{pvcharges}.  These additional chiral multiplets do not have any
superpotential couplings to the light chiral supermultiplets $Z^i$.
The condition \myref{gcond} is satisfied in the minimal supersymmetric
extension of the Standard Model (MSSM) and its extensions, as well as
in the hidden sectors~\cite{joel} of such extensions from all $Z_3$
orbifold compactifications of the heterotic string for which the full
spectrum is known.

\mysection{Supergravity}\label{sugra}
Supergravity is defined by the superpotential $W(Z^i)$, which is now an
arbitrary function of $Z$, the real K\"ahler potential
$K(Z^i,\Z^{\m})$ and the gauge kinetic function $f_{a b}(Z^i)$.  Here
I will assume that $f$ is diagonal:\footnote{The notation $\l X\r$
  stands for the lowest component (with the superspace coordinate
  $\theta = 0$) of the superfield $X$, with all fermion fields set to
  0.}
\beq f_{a b}(Z^i)=f(Z^i)\delta_{a b}\qquad \l f(Z^i)\r= f(z^i)
= x+iy,\eeq
which is the case for supergravity from the heterotic string.  To
obtain the one-loop effective (bosonic) Lagrangian~\cite{gj}, we
expand the action (covariantly) around a bosonic background, and
integrate over quantum fluctuations $h_{\mu\nu},\hat A^a_\mu,\hat z^i$
in the graviton, Yang-Mills and scalar fields, as well as fermions,
ghosts and an auxiliary field $\alpha$ that is used to implement the
gravitino gauge fixing.  For the bose sector we use smeared gauges,
defined by:
\bea \L_{gf} &=& -{\sqrt{g}\over 2}\(G_a G^a - G_\mu G^\mu\)\qquad
G_a = {1\over\sqrt{x}}\[\D_\mu\(x\hA_a^\mu\)
+ i K_{i\m}\(T_a^{\m}{\hz^i} - T_a^i{\hz^{\m}}\)\]\label{ymgauge}
\\ G_\mu &=& {1\over\sqrt{2}}\[\nabla^\nu{h_{\mu\nu}}
- {1\over 2}\nabla_\mu h^\nu_\nu - 2
\(\D_\mu\z^{\m}K_{i\m}{\hz^i}\hc\) + 
2x F^a_{\mu\nu}{\hA_a^\nu}\],\eea
while for the gravitino $\psi^\mu_\alpha$ we use an unsmeared gauge 
$$ G = - \gamma^\mu\(i\notD - \M\)\psi_\mu - 2K_{i\m}\[\(\notcD\z^{\m}
+i F^{\m}\)\chi^i + \(\notcD z^i + i F^i\)\chi^{\m}\] + \({x\over2}
\sigma^{\mu\nu}F^a_{\mu\nu} + {1\over x}\gamma_5D^a\)\lambda_a
=0,$$
\beq\delta(G) = \int d{\alpha}\;{\exp}\(i{\alpha}G\).\label{alpha}\eeq
The choice \myref{ymgauge} is the generalization to SUGRA of the
``supersymmetric gauge'' mentioned in \mysec{susyym}. We drop terms
that vanish by virtue of the tree equations of motion; much of this
can be done {\it a priori} by adding a judicious choice of such terms
to the inverse propagators.  With the above gauge fixing procedures,
the one-loop action takes the form
\beq S_1 = \half i\STr\[D^\mu D_\mu + H(g_{\mu\nu},F_{\mu\nu},z)\] 
+ T_-(g_{\mu\nu},F_{\mu\nu},z),\label{s1}\eeq
where $D_\mu(g_{\mu\nu},F_{\mu\nu},z)$ is a generalized covariant
derivative, $H$ a generalized squared mass matrix, and $T_-$ is the
helicity-odd fermion contribution.  The explicit expression for
$H$ is invariant under all the symmetries of the SUGRA theory.

\subsection{SUGRA with chiral matter}\label{sugchi}

In evaluating the one-loop quadratic divergences for supergravity
coupled to chiral matter~\cite{gj}, we use the trace of the
graviton equation of motion:
\beq \l g_{\mu\nu}{\del\L\over\del g_{\mu\nu}}\r_{\rm tree} =
{r\over2} - 2V + \D_\mu z^i\D^\mu\z^{\m}=0,\label{eom}\eeq
which is equivalent to a metric redefinition that restores the
Einstein term to canonical form in this order.  Here $V$ is the
scalar potential
\beq V= K_{i\m}F^i\bF^{\m} - 3M\bM,\qquad M = e^{K/2}W(z) = M_\psi,
\qquad K_{i\m} = {\pp^2K\over\pp z^i\pp\z^{\m}},\eeq
and $M$ is an auxiliary field\footnote{The normalization for $M$ used
  here differs by a factor $-{1\over3}$ from the the usual
  one~\cite{wb,bgg}.}  of the supergravity supermultiplet; its \vev\, is
the gravitino mass $M_\psi$.  The one-loop quadratic divergences
are determined by the sum of the supertraces from the gravity sector
and the chiral matter sector, which, using \myref{eom}, are given by
\bea \STr H_{\rm grav} &=&  - 14|M|^2
+ K_{i\m}\(4F^i\bF^{\m} - 3\D_\mu z^i\D^\mu\z^{\m}\),\label{strchi}  \\
\STr H_\chi &=& N_\chi\(2|M|^2 + K_{i\m}\D_\mu z^i\D^\mu\z^{\m}\)
- 2R_{i\m}\(F^i\bF^{\m} + \D_\mu z^i\D^\mu\z^{\m}\),\label{strgrav}\eea
where $R_{i\m}$ is the Ricci tensor derived from the K\"ahler metric
$K_{i\m}$, and $N_\chi$ is the number of chiral supermultiplets.  To
cancel the quadratic divergences we add the following PV
superfields~\cite{pvlett}:
\begin{MYitemize}{$\bullet$}
\item chiral superfields $Z^I$ with K\"ahler metric $K_{I\bM} = K_{i\m}$ 
and signature $\eta_I=-1$,
\item chiral superfields $\phi^\alpha$ with K\"ahler metric 
$K_{\alpha\bar\beta} = \del_{\alpha\bar\beta}e^{\alpha_\alpha K}$,
\item Abelian \uo\, vector fields $\Wa^n$ and \uo-charged chiral
fields $e^{\theta^n}$ which together form massive vector fields
by virtue of the superhiggs mechanism.
\end{MYitemize}
These give contributions
\bea \STr H^{P V}_\chi &=& N'_\chi\(2|M|^2 
+ K_{i\m}\D_\mu z^i\D^\mu\z^{\m}\) + 2\(R_{i\m}-\alpha K_{i\m}\)
\(F^i\bF^{\m} + \D_\mu z^i\D^\mu\z^{\m}\),\label{strchipv}\\
\STr H^{P V}_{\Wa} &=& N'_G\[K_{i\m}\(2F^i\bF^{\m} - \D_\mu
z^i\D^\mu\z^{\m}\) - 6|M|^2\], \label{strWpv} \eea
where
\beq N'_\chi = \sum_C\eta_{\phi_\chi^C},\quad\phi_\chi^C = Z^I,
\phi^\alpha,\theta_n, \qquad
N'_G = \sum_n\eta_{W^n_\alpha}, \qquad
\alpha = \sum_\beta\eta_{\phi^\beta}\alpha_\beta.\label{defs}\eeq
Cancellation of quadratic divergences is achieved with
\beq N'_\chi =3\alpha + 1 - N_\chi,\qquad N'_G = \alpha -2.
\label{conds}\eeq
Full cancellation of logarithmic divergences imposes an additional
constraint, giving~\cite{pvcan} 
\beq N'_\chi = -29 - N_\chi,\qquad
N'_G = -12,\qquad \alpha = - 10.\label{conds1}\eeq
It also requires the introduction of additional PV chiral superfields
$Y_I$ with K\"ahler metric $K_Y^{I\bM} = K^{i\m}$, $K^{i\m}K_{j\m} =
\del^i_j$, as well as several copies of the $Z^I$ with alternating
signatures. All of these are included in the definition of $N'_\chi$
in \myref{defs}.

\subsection{SUGRA with YM and chiral matter}\label{minym}

Now we add to the theory Yang-Mills superfields with canonical kinetic
energy terms:
\beq f(z^i)=x+i y = g^{-2} - i{\theta\over8\pi}= {\rm constant}.\eeq
If the chiral multiplets have gauge couplings as in \myref{charges},
the potential acquires a D-term
\beq  \Del V = \half x D^a D_a,\qquad D_a = K_i(T_a z)^i 
= K_{\m}(T^T_a\z)^{\m},\qquad K_i = {\pp K\over\pp z_i}, 
\quad K_{\m} = {\pp K\over\pp z_{\m}}.\eeq
The PV superfields $Z^I$ now transform as in \myref{pvcharges}, and
the supertraces \myref{strchi}, \myref{strchipv} and \myref{strWpv} get the
additional contributions~\cite{gjs}
\bea \Del\STr H_\chi &=& -  N_\chi x D^a D_a +
2D_a\[\Gamma^i_{i j}(T^a z)^j + (T^a)^i_i\],\\
\Del\STr H^{P V}_\chi &=& \(2\alpha - N'_\chi\)x D^a D_a -
2D_a\[\Gamma^i_{i j}(T^a z)^j + (T^a)^i_i\],\\
\Del\STr H^{P V}_{\Wa} &=& N'_G x D_a D^a,\eea
where $\Gamma^i_{j k}$ is the ``affine connection'' associated with
the K\"ahler metric.  In addition, there is an off-diagonal mass term
connecting the gaugino to the auxiliary field $\alpha$ introduced in
\myref{alpha}:
\beq M_{\alpha\lambda^a}= -\sqrt{x}\(D_a +{1\over2}F_a^{\mu\nu}
\sigma_{\mu\nu}\)
= - \sqrt{x}\[D_a + {1\over2}\(\beta F_a^{\mu\nu} + i\gamma\gamma_5
 \tF_a^{\mu\nu}\)\sigma_{\mu\nu}\],\qquad
\beta + \gamma = 1.\label{gamma2}\eeq
The second equality in \myref{gamma2} follows from the properties of
Dirac matrices; it illustrates the ambiguity in defining $\gamma_5$
that is present in any regularization procedure.  The
``supersymmetric'' choice is
\beq \beta = 1,\qquad \gamma = 0.\eeq
With this choice \myref{gamma2} matches off-diagonal squared masses
that couple $h_{\mu\nu}$ to $\hat A_\mu$ and the graviton ghost
$c_\mu$ to the YM ghosts $c_a$, and it allows for BRST invariant PV
regularization.  The Yang-Mills sector gives a contribution
\bea \STr H_{\rm Y M} &=& 
(1+N_G)x D_a D^a + \half x
  F_{\mu\nu}F^{\mu\nu}\ddd + N_G\[K_{i\m}\(2F^i\bF^{\m} - \D_\mu
z^i\D^\mu\z^{\m}\) - 6|M|^2\],\label{strW}\eea
and \myref{strgrav} gets an additional contribution
\bea\Del\STr H_{\rm grav} &=& 2x D_a D^a
 - \half x F_{\mu\nu}F^{\mu\nu}.\eea
 The terms containing the YM field strength cancel, and all UV
 divergences can be canceled~\cite{pvlett} provided $N'_G$
in \myref{conds} and \myref{conds1} is shifted by the amount
\beq \Del N'_G = -N_G.\eeq
Full cancellation of logarithmic divergences~\cite{pvcan} requires
including the chiral superfields $Y_I$, with K\"ahler metric as in
\mysec{sugchi} and gauge charges as in \myref{pvcharges}, the chiral
superfields $\varphi^a$ in the adjoint representation of the gauge
group that were introduced in \mysec{susyym}, as well as additional
copies of these and other chiral superfields, such that, in
particular, \myref{gcond} is satisfied.

\subsection{Including the dilaton}\label{dilsug}
Finally, we include a nontrivial gauge kinetic function:
\beq f(z^i)= \l f(Z^i)\r = x(z^i) +i y(z^i),\qquad\myvev{x(z^i) +i
  y(z^i)} = g^{-2} - i{\theta\over8\pi},\qquad f_i = \pp_i f\ne0.\eeq
This introduces~\cite{gjs} an additional off-diagonal mass term that
mixes gauginos with the fermion superpartner of the dilaton $f(z^i)$:
\beq\Del M_{\chi^i\lambda^a}= - i{f_i\over2\sqrt{x}}\[D_a + 
\(\beta F_a^{\mu\nu} + i\gamma\gamma_5\tF_a^{\mu\nu}\)\sigma_{\mu\nu}\],
\qquad \beta + \gamma = 1.\eeq
In this case the ``supersymmetric'' choice is
\beq \beta = \gamma = \half,\label{susychoice}\eeq 
which matches a squared mass term that couples the dilaton to
the Yang-Mills fields, and BRST invariant PV regularization
is again possible.  The YM field strength terms vanish identically
in the  squared masses, {\it e.g.},
\beq\left|\Del M_{\chi^i\lambda^a}\r^2 = {f_i\f^i\over 4x}D_a D^a,
\qquad \f^i = K^{i\m}\f_{m},\eeq
and the new contribution to the chiral multiplet supertrace is
\beq \Del\STr H_\chi = {f_i\f^i\over4x}D^a D_a.\label{dilchi}\eeq
There is also an additional term in the gaugino connection: 
\beq\Del A^\mu_{\lambda} = - {\pp^\mu y\over2x}\(i\delta\gamma_5
- \epsilon{\epsilon^{\tau\nu\rho\sigma}\over24}\gamma_\tau
\gamma_\nu\gamma_\rho\gamma_\sigma\), \qquad 
\del + \epsilon = 1.\label{newcon}\eeq
We choose: 
\beq\del=0,\qquad\epsilon = 1,\qquad\Del A^\lambda_\mu =
2x h^{\nu\rho\sigma}\gamma_{[\mu}\gamma_\nu\gamma_\rho\gamma_{\sigma]},
\label{dilchoice}\eeq
where $h_{\mu\nu\rho}$ is the three-form, dual to the axion $y$, that
comes from compactification of the ten dimensional supergravity limit
of the heterotic string.  With the choice \myref{dilchoice} the
connection \myref{newcon} is a vector current.  There is no associated
anomaly, the QCD vacuum angle $\theta$ is not renormalized, in
agreement with earlier results~\cite{sv}, and the modified linearity
condition is respected~\cite{gt} at one-loop order in the dual linear
multiplet formulation for the dilaton supermultiplet.  Specifically,
in the effective supergravity theory from the WCHS, the dilaton
supermultiplet $f(Z) = S(s,\chi^s_\alpha,F^s)$ is dual to a (modified)
linear supermultiplet $L(\ell,\chi_\alpha^\ell,b_{\mu\nu})$, where
$b_{\mu\nu}$ is a two-form whose curl is the three-form
$h_{\mu\nu\rho}$ in \myref{dilchoice}.  With the choices
\myref{susychoice} and \myref{dilchoice}, the new contributions to the
YM supertrace are
\bea \Del\STr H_{\rm Y M} &=& -{f_i\f^i\over4x}D_a D^a
- {N_G\over2x^2}\[f_i\f_{\m}F^i\bF^{\m} + \(\pp_\mu x\pp^\mu x 
+\pp_\mu y\pp^\mu y\)\].\label{dilYM}\eea
The D-terms in \myref{dilchi} and \myref{dilYM} cancel, and we
obtain an overall contribution
\bea \Del\STr\(H_\chi + H_{\rm Y M}\) &=& 
- {N_G\over2x^2}\[f_i\f_{\m}F^i\bF^{\m} + \(\pp_\mu x\pp^\mu x + 
\pp_\mu y\pp^\mu y\)\]\label{diltot}.\eea
This contribution can be canceled by
adding~\cite{pvlett} chiral PV multiplets $\pi^\alpha$ with K\"ahler metric
$K(\pi,\bar\pi) = (f + \f)|\pi|^2$ and/or by coupling~\cite{pvdil} some
Abelian gauge PV multiplets to the dilaton, that is, by setting
$f_{\Wa^n} = e_n f(Z)$. Cancellation of \myref{diltot} requires
\beq N_\pi - e = N_G\qquad N_\pi={\sum_\alpha}\eta_{\pi^\alpha}
\qquad e={\sum_n}\eta^{n}e_n.\eeq
Cancellation of logarithmic divergences requires~\cite{pvdil} the
second mechanism:
\beq N_\pi=0\qquad e= -N_G.\eeq
Large PV masses for the chiral superfields
$Z^I,Y_I,\varphi^a,\phi^\alpha$, as well as those needed to assure
that the condition \myref{gcond} is satisfied, are generated by
including gauge invariant bilinears of these superfields in the 
superpotential, and large PV masses for the ($\Wa^n,\theta^n$) arise
from the Abelian superhiggs mechanism.  The squared cut-off in the UV
divergent terms are replaced by the relevant squared PV masses, and
one obtains an expression of the form
\beq \L_{\small\mbox{tree + 1-loop}} =
\L_{\rm{tree}}(g^R_{\mu\nu},K^R,g_a^R) + \mbox{operators dim}\ge
6.\eeq
All of the higher dimension terms that cannot be absorbed into
renormalizations (denoted by the superscript $R$) are associated with
UV logarithmic divergences.

\mysection{Two Applications}\label{loop} 
In this section I will describe applications to particle phenomenology
of the regularization procedure described above.  Both cases illustrate
the sensitivity of the scalar potential to the choice of PV masses,
which cannot be completely fixed by the requirement of UV finiteness.

\subsection{Taming large quadratic divergences} 
It has been pointed out~\cite{ckn,clm} that the loop suppression parameter
\beq \ep = {1\over16\pi^2}\eeq 
may be compensated by large coefficients, leading to significant
effects from loop corrections. Specifically, once \myref{eom} is
imposed, the quadratically divergent correction to the scalar
potential includes the terms:
\beq V_Q = {\half}\ep\Lambda^2\STr H_{\rm nonderiv} \ni
\ep\Lambda^2\[|M|^2\(N_\chi - 3N_G - 7\) - N_G M^2_\lambda
- R_{i\m}F^i\bF^{\m}\].\label{vquad}\eeq
Typical WCHS orbifold compactifications have many more chiral
multiplets than gauge multiplets:
$N_\chi\gappeq 300,\quad N_G \lappeq 65$, so since in many
gravity mediated supersymmetry-breaking scenarios the gaugino
mass $M_\lambda$ is much smaller than the the gravitino mass:
\beq M_\lambda^2 = {1\over4}f_i\f^i M^2\ll M^2,\eeq
the first term in \myref{vquad} suggests the possibility of a
significant {\it positive} contribution to the vacuum
energy~\cite{ckn}, perhaps curing the problems with classes of models
that have negative vacuum energy at tree level.  However, in the
regulated theory \myref{vquad} is replaced by
\beq  V_Q\to\ep\[|M|^2\(N_\chi\Lambda^2_\chi - 3N_G\Lambda^2_G
- 7\Lambda^2_{\rm  grav}\) - N_G M^2_\lambda\Lambda'^2_G
- R_{i\m}F^i\bF^{\m}\Lambda'^2_\chi\] + \cdots,\label{vreg}\eeq
where the ellipsis indicates finite terms proportional to $M^2_{P V}$ 
such that the one-loop quadratically divergent corrections are
completely absorbed into renormalizations:
\beq\L_Q = \L_{\rm tree}(g^R_{\mu\nu},K^R) - 
\L_{\rm tree}(g_{\mu\nu},K) + O(\ep^2),\qquad
K^R = K + \ep{\sum_A}\Lambda^2_A.\label{lquad}\eeq
The effective squared cut-offs $\Lambda^2_A$ in \myref{vreg} and
\myref{lquad} are determined by the PV masses:
\beq \Lambda^2_A = C_A\(\eta^i M^2_i\ln M^2_i\)_A,\qquad
\({\sum_i}\eta_i M^2_i\)_A = 0,\eeq
where $C_A$ is a constant.  Several remarks are in
order~\cite{quaddiv}.
\begin{itemize}
\item The sign of $\Lambda^2_A$ is indeterminate~\cite{sigma} if 
there are five or more terms in the sum, which is generally required
to eliminate all the UV divergences of SUGRA.
\item If $N_\chi\sim\ep^{-1}$ one has to sum the leading
  $(\ep\Lambda^2)^n$ terms~\cite{quaddiv}. 
\item Supersymmetry dictates that the higher order terms 
complete the Lagrangian $\L_{\rm tree}(g^R_{\mu\nu},K^R)$ with $K_R$
given by \myref{lquad}.
\end{itemize}
So, for example, if the $M^2_i$ are field independent constants, we
just get
\beq V_Q = e^{K+\Del K}\(W_i K^{i\m}\W_{\m}- |W|^2\) + \half x D^a D_a,
\qquad W_i = {\pp\over\pp z^i}W = - e^{-K/2}K_{i\m}\bF^{\m}.\eeq
If, in addition, supersymmetry is broken only by F-terms, $\myvev{D_a}
= 0$, the vacuum energy is just multiplied by a positive constant.

It has also been pointed out~\cite{clm} that the last term in
\myref{vquad} or \myref{vreg} can be significant because it involves a
sum over all the chiral supermultiplets.  The K\"ahler potential for the
untwisted sector from orbifold compactification of the heterotic
string is not known beyond leading (quadratic) order, and could
include terms that induce flavor changing neutral current (FCNC)
effects in the observable sector.  Experimental limits on these
effects therefore imply restrictions on the K\"ahler potential.  A
sufficient condition~\cite{quaddiv} for a ``safe'' K\"ahler potential
is the presence of isometries of the K\"ahler geometry.  For example,
the K\"ahler potential for an untwisted sector $n$ from orbifold
compactification takes the form
\beq K^n = - \ln(T^n + \bar T^{\bar n}
    -{\sum_{A=1}^{N_n}} |\Phi^A_n|^2), \eeq 
which has an $SU(N_n + 1,1)$ symmetry that is necessarily also 
a symmetry of the Ricci tensor:
\beq R^n_{i\m} = (N_n + 2)K^n_{i\m}.\eeq
Alternatively the suppression of FCNC effects can by achieved
through a judicious choice of PV masses~\cite{quaddiv}.

\subsection{Anomaly mediated SUSY breaking}
One-loop contributions to soft supersymmetry breaking parameters for
the superpartners of the Standard Model particles can be important,
particularly in models where they are suppressed at tree level.  If
they arise only through loop effects, the mechanism for supersymmetry
is referred to as ``anomaly mediation''.

The one-loop contribution to gaugino masses $m_a$ is 
independent of Planck-scale physics, and is completely determined by
the properties of the effective low energy (sub-Planck scale)
theory.  The result is~\cite{gny,bmp,gn}
\beq \Del m_a(\mu) = -3\beta_a(\mu)M - {g^2(\mu)\over14\pi^2}F^j\[
\(C_a - C^M_a\)K_j + 2\sum_i C^i_a\pp_j\ln K_{i\ibar}\].\eeq
The term proportional to the $\beta$-function \myref{beta} is
related~\cite{glmr,rs} to the conformal anomaly, in that it arises
from the running of the coupling constant from the Planck scale to the
scale $\mu$, and has been shown to be exact to all orders in
perturbation theory.

Writing the superpotential in the form
\beq W(Z) = \sum_{i j k}W_{i j k}Z^i Z^j Z^k + 
\sum_{i j}\mu_{i j}Z^i Z^j + O(Z^4), \eeq
supersymmetry breaking generates so-called $A$ and $B$ terms in
the scalar potential that are, respectively, cubic and quadratic
in the scalar fields $z^i$:
\beq V\ni {1\over6}{\sum_{i j k}}A_{i j k}W_{i j k}z^i z^j z^k + 
{\half}{\sum_{i j}}B_{i j}\mu_{i j}z^i z^j\hc\eeq
Neglecting small flavor mixing in the anomalous dimension matrix
\myref{gamma}, the one-loop contributions to the parameters $A$ and
$B$ are~\cite{gn,bgn}
\bea \Del A_{i j k}(\mu) &=& \(\gamma_i + \gamma_j +
\gamma_k\)_\mu M + a_{i j k}\ln(M_{P V}/\mu),\\
\Del B_{i j}(\mu) &=& \(\gamma_i + \gamma_j\)_\mu M
+ b_{i j}\ln(M_{P V}/\mu).\eea
The first term in each expression is the conformal anomaly
contribution~\cite{rs,glmr}, again valid to all orders in perturbation
theory.  The (field-dependent) parameters $a$ and $b$ vanish if there
are no tree-level soft terms in the observable sector.

In contrast to the above, the supersymmetry-breaking (``soft'')
scalar squared masses $m^2_i$ are strongly dependent on
Planck scale physics~\cite{gn}:
\beq \Del m^2_i = 9\gamma_i|M|^2 + \nu_i(m^{P V}_{\rm soft}) + \mu_i,
\label{softm2}\eeq
where the last term vanishes if there is no tree-level SUSY breaking
in the observable sector.  The first, ``conformal anomaly'', term was
not found in earlier analyses~\cite{glmr,rs}; they found instead a
universal two-loop contribution proportional to the derivative of the
anomalous dimension matrix $\gamma$. The second term vanishes only if
the tree-level {\it Pauli-Villars} soft squared masses vanish, which
is generally not the case.  In the ``sequestered sector'' model of
Randall and Sundrum~\cite{rs} the first term is exactly canceled by
the second; this requires\footnote{It was noted in~\cite{pr} that this
  result rests on the assumption that $\myvev{F^i K_i}$ is
  negligible.}  a very special form of the hidden sector scalar
potential, as well as of PV masses. The spurion analysis~\cite{glmr}
missed the second term in \myref{softm2} because of an assumption of
holomorphicity that is not borne out by the explicit PV
calculation~\cite{gn}.

The sensitivity of soft scalar masses to Planck scale physics can
easily be understood in the framework of PV regularization.
Superpotential and gauge couplings of light chiral superfields are
regulated by the PV fields $\Phi^A= Z^I,Y_I,\varphi^a$, which obtain
large PV masses through gauge invariant superpotential couplings to
other fields $\Phi'_A$:
\beq W_{P V}\ni \mu_A\Phi^A\Phi'_A,\qquad m^2_A = m^2_{A'} =
e^K K^{A\A}_\Phi K^{\Phi'}_{B\bar B}|\mu_A|^2.\label{WPV}\eeq
The finiteness requirement constrains the K\"ahler metric for
$\Phi^A$, but not for $\Phi'_A$, since they need not have any
couplings to light sector fields, except for electromagnetic
couplings if they carry gauge charges.  Since {\it all} 
gauge-charged PV fields contribute to the $\beta$-functions
\myref{beta}, the PV loop contribution to the gaugino masses
is uniquely fixed. On the other hand the fields $\Phi'_A$ need
not have any superpotential couplings to the light fields, so
the constraint that the UV divergence associated with the anomalous
dimension matrix $\gamma$ vanishes places no restriction on their K\"ahler
metric, and no restriction on the corresponding PV masses, so
$\Del m_i^2$ is undetermined.  

There is a parallel situation concerning the K\"ahler chiral
and conformal anomalies associated, respectively, with linear
and logarithmic UV divergences.  Supergravity is classically
invariant~\cite{bz} under K\"ahler transformations
\bea K&\to& K + F(Z) + \bF(\Z),\qquad W\to e^{-F}W, \nnn \chi&\to&
e^{i\im F/2}\chi,\qquad \lambda\to e^{i\im F/2}\lambda,\qquad
\psi\to e^{i\im F/2}\psi,\label{kahler}\eea
which are anomalous at the quantum level.  The chiral anomaly of the
Yang-Mills Lagrangian associated with the phase transformation on the
fermions in \myref{kahler} forms an F-term superfield component
together with the conformal anomaly associated with the
$\beta$-functions; this operator is completely fixed by the
requirements of UV finiteness and supersymmetry, and is independent of
Planck scale physics or the regularization procedure.  By contrast,
the conformal anomaly associated with the $\gamma$-functions is a
D-term superfield component, with no chiral counterpart, and 
depends on the details of the regularization procedure, which in
turn should parametrize Planck scale physics. The WCHS is 
perturbatively invariant under all gauge transformations, as
well as a group of transformations on the chiral superfields
$Z = T^n,\Phi^i$, 
\beq T^n\to f(T^n), \qquad \Phi^i\to g(q^i_n,T^n)\Phi^i,\eeq
called T-duality, that effects a K\"ahler transformation
\myref{kahler} with $F = F(T^n)$, with the fields $T^n$ known as
``K\"ahler moduli'', and $q^i_n$ the ``modular weights'' of $\Phi^i$.
However the effective quantum field theory is anomalous under
T-duality.  The regularized theory is anomaly free if PV mass terms
respect the classical symmetries. This is not possible in the case of
T-duality, or for an anomalous Abelian symmetry, \ux, that is present
in almost all realistic theories that break part of the gauge symmetry
at the string scale by Wilson loops.  For example, the PV superfield
$\phi^\beta$ gives a contribution to the quadratically divergent
one-loop Lagrangian
\beq (\L_Q)_{\phi^\beta}\propto\STr H_{\phi^\beta} \ni \(1 -
2\alpha^\beta\) \(K_{i\m}\D_\mu z^i\D^\mu\z^{\m} - x D^a D_a\)+ 2D_a
q^\beta_X.\label{phi}\eeq
To obtain an invariant mass, $\phi^\beta$ must have a superpotential
coupling to another field $\phi^\gamma$ with
\beq \alpha^\beta + \alpha^\gamma = 1,\qquad q_X^\beta + q_X^\gamma =
0,\eeq
such that the contribution from $\phi^\gamma$ exactly cancels
\myref{phi}. One could instead restore T-duality by making the mass
parameters in \myref{WPV} field-dependent: $\mu\to \mu(T^n)$; this
would be interpreted as a threshold correction~\cite{thresh}.  However
such corrections are known be absent~\cite{ant} in, for example, $Z_3$
and $Z_7$ orbifold compactifications.

\mysection{Anomalies and anomaly cancellation in supergravity} 
It has long been known how to cancel the T-duality~\cite{modular} and
\ux\,~\cite{ux} anomalies involving Yang-Mills field strength
bilinears.  The full anomaly structure of PV regulated supergravity
has been determined only recently~\cite{anom}; its detailed form, and
therefore the possibility of anomaly cancellation, depends on the
choice of PV couplings.  It was recently shown~\cite{ss} that for
specific $Z_3$ and $Z_7$ compactifications, with no Wilson lines and
therefore no anomalous \ux, the {\it string theory} anomaly is
completely canceled by the four dimensional version of the
Green-Schwarz mechanism~\cite{gs}. If PV regularization can be a
faithful parameterization of the higher string and Kaluza-Klein modes
that render the full theory finite, there should be a choice that
realizes this result at the field theory level; determining this
prescription could in turn restrict the loop corrections to the scalar
potential discussed in \mysec{loop}.

\subsection{Anomalous YM couplings, their cancellation and two applications
to phenomenology}\label{anomYM}
Under T-duality and \ux, the shift in the YM Lagrangian is given, in
the K\"ahler \uo\, superspace formulation~\cite{bgg} of SUGRA, by the
expression
\beq\Del\L_{\small\mbox{YM loop}}={1\over8}{\sum_a}\superint{E\over R}
\[{\sum_n}c_a^n H(T^n) + c_a\Lambda_X\](\WaWa)_a\hc,\label{delLYM}\eeq
where
\beq c^n_a = {1\over8\pi^2}\[C_a - \sum_i C_a^i(1 - 2q^i_n)\],\qquad
c_{a\ne X} = \displaystyle{1\over4\pi^2}{\Tr T^2_a T_X},\quad c_X =
{1\over12\pi^2}{\Tr T_X^3}.\eeq
The anomaly is canceled by a 4-dimensional version of the
Green-Schwartz (GS) mechanism; the dilaton is no longer invariant
under these transformations:
\beq \Del S = -b H(T) - c\Lambda_X.\eeq
Then the variation \myref{delLYM} is canceled by the variation of the
tree Lagrangian:
\beq \L_{\mbox{YM tree}} = {1\over8}\superint{E\over R}S\sum_a(\WaWa)_a\hc,
\qquad \Del\L_{\mbox{YM tree}} = -\Del\L_{\small\mbox{YM loop}}.\eeq
To make the theory fully invariant, the dilaton K\"ahler potential
$K(S+\S)$ is replaced by 
$$K[S+\S + V(T,\T) + c\vx],$$ 
where $\vx$ is the \ux\, vector superfield:
\beq\Del\vx = \Lambda_X + \bar\Lambda_X,\eeq
and the function $V(T,\T)$ satisfies
\beq\Del V = H + \bar H.\eeq
The full $T$-dependence of $V$ is determined by matching~\cite{gt} string
and field theory calculations of the $\im t F\cdot\tF$ vertex:
\beq V(T,\T) = - \sum_n\ln(T^n + \T^{\n}).\eeq
Anomaly cancellation requires
\beq c_a = c = {\Tr T_X\over96\pi^2}\qquad \forall\quad a,\qquad
c_a^n = b\qquad \forall\quad a,n\label{gscond}\eeq 
for compactifications with no threshold corrections, such as $Z_3$ and
$Z_7$ orbifolds.  For those with string loop threshold corrections of
the form
\beq \L_{\rm th} = \sum_n{b_a^n\over8}\superint{E\over R}f(T^n)\sum_a 
(\WaWa)_a\hc,\qquad \Del f(T^n) = H(T^n),\eeq
the second condition in \myref{gscond} is replaced by
\beq  b_a^n = b - c_a^n.\eeq
Note that the one-loop calculation yields a supersymmetric anomaly;
the higher loop corrections to the $\beta$-function are encoded in 
the PV cut-off demanded by supersymmetry~\cite{gt,fmtv,bg}; for example
\beq\Lambda_G^2 = e^{K/3} = \[16(\re s)\prod_n(\re t^n)\]^{-{1\over3}} = 
g^{-{4\over3}}R_{\rm comp}^{-2} = g^{-{4\over3}}\Lambda_{\rm comp}^{2},
\eeq
where the subscript ``comp'' refers to the compactification radius/scale.

These results have two important applications to phenomenology:
\begin{MYitemize}{$\bullet$}
\item Matching the coefficient of $F^a_{\mu\nu}\cdot F^{\mu\nu}_a$ to the
two-loop RGE invariant~\cite{sv} of supersymmetric Yang-Mills theories
fixes~\cite{gt,kl} the gauge unification scale; this gives in the 
$\overline{MS}$ scheme
\beq  \mu_{\rm unif}^2={m^2_{\rm string}\over2e} = 
{g^2m_{\rm Planck}^2\over2e}\sim 2\times 10^{17}{\rm GeV}.\eeq
This is an order of magnitude lower than what is obtained by
extrapolating from low energy data in the context of the minimal
supersymmetric extension of the Standard Model, but in effective
theories from superstrings one expects heavy states that are
vector-like under the Standard Model gauge group, as well as
corrections to the dilaton K\"ahler potential from string nonperturbative
effects and/or field theory loop effects.

\item The effective Lagrangian $\L_{\rm eff}(U_a,\Pi^i)$ for gaugino
  condensates $U_a\simeq(\Wc_a\Wa^a)_{\rm hid}$ and matter condensates
  $\Pi^i\simeq\prod_A(\phi^A)^{n^i_A}_{\rm hid}$ in a strongly coupled
  hidden sector can be constructed by matching~\cite{bgw} the
  anomalies of $\L_{\rm eff}$ to \myref{delLYM}, thus providing a
  mechanism for supersymmetry breaking.
\end{MYitemize}

\subsection{Full anomaly cancellation?}

The linear divergences of supergravity can be canceled by the PV
fields introduced in \mysec{sugra}, except for some from
nonrenormalizable terms in the $\psi,\lambda$ connections.
The residual chiral anomalies associated with the latter terms
form supersymmetric (F-term) anomalies together with residual conformal
anomalies proportional to total divergences, {\it provided} the cut-off
is field-dependent:
\beq \Lambda(Z)= e^{K/2}\Lambda_0,\qquad\Lambda_0\to\infty.\eeq
The resulting effective theory is fully finite, with the remainder of
the anomalies arising from a subset of chiral PV superfields with
noninvariant masses \myref{WPV}, that can be chosen to have a simple
K\"ahler metric.  The total anomaly in the regulated theory is then
given by~\cite{anom,bg3}
\bea\Del\L &=&{1\over8\pi^2}\[\Tr\eta\Phi(T^n,\Lambda_X)
\({1\over3}\Omega_W + \Omega_{\rm Y M} + \cdots\) + H(T)\(
\Omega_W + \cdots\)\]\hc\eee
\Del\L_{\mbox{YM loop}} + {1\over8}\(\superint{E\over R}\[\sum_n c_n H(T^n)
+ c\Lambda_X\]W^{\alpha\beta\gamma}W_{\alpha\beta\gamma}\hc\)
+ \cdots\label{DelL} \eea
The first term in the first line of \myref{DelL} comes from the
noninvariant PV masses, and the second from the variation of the
effective cut-off:
\beq \Del\ln M_{P V} = \Phi, \qquad \Del K = H + \H.\eeq
$\Omega_{\rm YM}$ and $\Omega_W$ are Chern-Simons superfields whose
chiral projections are, respectively, the Yang-Mills and curvature superfield
strengths:
\beq\chiproj\Omega_{YM} = \sum_{a\ne X}\Wc_a\Wa^a, 
\qquad\chiproj\Omega_W = W^{\alpha\beta\gamma}W_{\alpha\beta\gamma}.\eeq
The terms proportional to $\Omega_{\rm YM}$ are just those found in 
\myref{delLYM}, and can be canceled as in \mysec{anomYM}. In the case
where threshold corrections are present, these can be included by an
appropriate $T$-dependence in the PV masses. The constants $c_n$ and $c$
are determined by the requirement that on-shell quadratic UV divergences
vanish; $c$ is given by \myref{gscond}, and
\beq c_n = {1\over192\pi^2}\(2\sum_A q^A_n - N_\chi + N_G -21\).\eeq
We have checked~\cite{bg3} for specific $Z_3$ and $Z_7$ orbifolds, with~\cite{fiqs}
and without~\cite{ss} Wilson lines, that $c_n = b$, so the term
proportional to $\Omega_W$ can also be canceled by the GS mechanism,
provided the tree Lagrangian contains a term
\beq L_{\rm tree}\ni-\superint E(S+\S)\Omega_W,\eeq
which is indeed present in effective supergravity from the heterotic
string.  The ellipsis in the second parentheses in 
\myref{DelL} represents ``D-term'' anomalies from additional logarithmic
divergences of the form
\beq\L_{\small\mbox{1 loop}}\ni \pp_\mu O^\mu\ln\Lambda^2;\eeq
these have not yet been completely determined.  The ellipsis in the
first parenthesis in \myref{DelL} represents terms nonlinear in the
parameters of anomalous transformations.  Their coefficients depend on
the detailed choice of the PV K\"ahler potential, and therefore of the
PV masses.  The challenge is to find a choice that mimics the string
result~\cite{ss}.  It may also be the case that full cancellation of
the anomalies requires constraints on the K\"ahler potential for the
twisted sector, analogous to the constraint \myref{gcond} on gauge
charges that is required for cancellation of UV divergences.
Resolving these questions would have important implications for the
phenomenological issues discussed in \mysec{loop}.

\mysection{Afterword}
Although Raymond and I have never written a paper together, we did
have one very successful collaboration.  Of the 51 students at the
1981 Les Houches summer school that we co-directed, at least 38 (some
at this meeting) are still active in particle physics, and many are
leaders in the field, not just in terms of scientific productivity,
but also in terms of service to the scientific community.\newline
\centerline{\large\it Bonne Aniversaire Raymond!}
\begin{figure}[h]%[htb!]
\centering%
\scalebox{0.5}
{\includegraphics{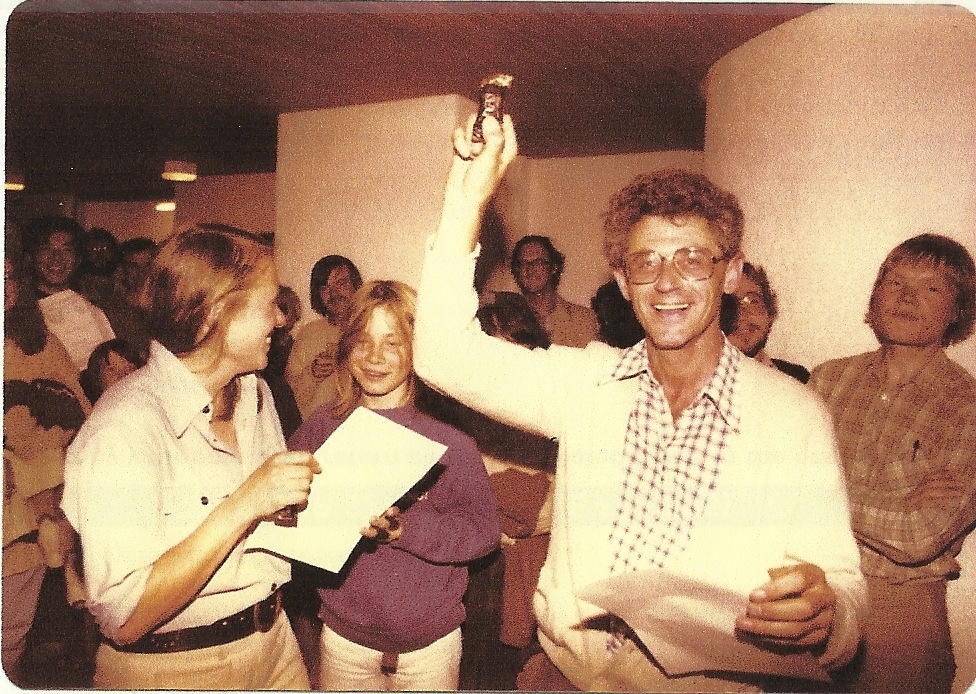}}
%\caption{An Example Figure}
%\label{fig:FigureExample}
\end{figure}
\section*{Acknowledgment}
This work was supported in part by the Director, Office of Science,
Office of High Energy and Nuclear Physics, Division of High Energy
Physics of the U.S. Department of Energy under Contract
DE-AC02-05CH11231, in part by the National Science Foundation under
grant PHY-0457315.

\end{document}